\documentclass[journal]{IEEEtranTIE}
\usepackage{graphicx}
\usepackage{cite}
\usepackage{picinpar}
\usepackage{amsmath}
\usepackage{url}
\usepackage{flushend}
\usepackage[utf8]{inputenc}
\usepackage{colortbl}
\usepackage{soul}
\usepackage{array}
\usepackage{makecell}
\usepackage{multirow}
\usepackage{booktabs}
\usepackage{pifont}
\usepackage{color}
\usepackage{alltt}
\usepackage[hidelinks]{hyperref}
\usepackage{enumerate}
\usepackage{siunitx}
\usepackage{breakurl}
\usepackage{epstopdf}
\usepackage{pbox}
\usepackage{tikz}
\usepackage{circuitikz}
\usetikzlibrary{shapes.geometric, arrows}
\usepackage{xcolor}
\definecolor{darkgreen}{rgb}{0,0.5,0} 
\usepackage{subcaption}
\usepackage{pgfplots}
\pgfplotsset{compat=1.18}

\tikzstyle{startstop} = [rectangle, rounded corners, text centered, draw=black]
\tikzstyle{process} = [rectangle, text centered, draw=black]
\tikzstyle{arrow} = [thick,->,>=stealth]

\begin{document}

\title{
    Winding the Field, Not the Wire: Overlaying Magnetics to Improve Current Dynamics in Modular Power Electronics
}

\author{
	\vskip 1em
	Jinshui Zhang, \emph{Student Member},
	Boshuo Wang, \emph{Senior Member},
    Hunter Cullina,
    Emmy Nelson Madore, 
    Maya Clinton,
	Angel V Peterchev, \emph{Fellow},
	and Stefan M Goetz, \emph{Member}
	\thanks{
	}
}

\maketitle
\begin{abstract}
    Modular power electronics scale power capacity by distributing it across cascaded modules.
    State-of-the-art modular multilevel converters further introduce dynamic inter-module switched-capacitor exchange and balancing through parallelization. 
    In larger converters, however, this mode needs inter-module inductance to control the equilibration currents. 
    Specifically, differential-mode inductance limits these currents, but existing solutions require conductor re-routing and introduce parasitic resistance.
    This work presents a non-intrusive resistance-free differential-mode inductance solution: a figure-of-eight-shaped magnetic core that guides the magnetic flux path around existing conductors, such as high-current bus bars.
    This design adds differential-mode inductance without modifying the conductors.
    We fabricated a core that provides a 1.52~µH differential-mode inductance while suppressing the common-mode inductance to only 7\% of the DM inductance.
    The core reduces the inter-module current surge from 20.46~A to 7.76~A. The original conductors remain intact.
\end{abstract}



\section{Introduction}

High-power applications such as renewable energy systems, large drives, and pulse generators increasingly rely on modular architectures to scale power capacity \cite{bohler2022ultra}.
In these architectures, modules connect in parallel to equilibrate module voltages and exchange energy, or in series to achieve higher voltages \cite{bohler2022ultra}.
Hybrid switched-capacitor\,/\,cascaded-bridge circuits combine both features: modules in series synthesize a staircase output voltage on the inter-module common mode (CM); the parallel operation can balance modules on the inter-module differential mode (DM)\cite{lizana2018modular, li2020reduced}. The losses are proportional to difference of the squared module voltage and are smaller than for other balancing methods if controlled properly.

\begin{figure}
    \centering
        \includegraphics{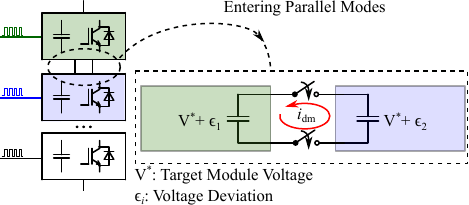}
    \caption{
    Illustration of dynamic inter-module parallelization.
    }
    \label{fig:scenario_of_parallel_choke}
\end{figure}

However, in larger converters with low intrinsic inter-module impedance, this feature of dynamic parallelization requires some inductance to control the equilibration currents. 
Surges with small inter-module impedance can stress the transistors and reduce system reliability (Fig. \ref{fig:scenario_of_parallel_choke}) \cite{goetz2014modular,zhang2024analytical}. 
With sufficient inductance, the interconnection can even operate as an inter-module buck--boost switched-capacitor converter and allow free voltage control between modules \cite{lizana2018modular}.

This inductance does not need to apply to all inter-module currents but only DM currents, which represent local inter-module energy exchange, without affecting the CM global cascaded-bridge output. In larger converters, the magnetics of such an inductor with high DM inductance but low CM value can be notably smaller, as it only needs to manage the flux densities of the smaller local balancing currents.

Existing methods for inter-module DM inductance include discrete inductors \cite{lu2024hybrid}, flexible wire couplers with magnetic cores \cite{li2020reduced}, helix busbar designs \cite{mao2017balancing}, and PCB-integrated solutions \cite{zhang2024analytical}. 
All require the conductors to be extended and routed to a coupling structure. 
Such rerouting redesigns the interconnection layout and lengthens the conductors, which increases their resistance as well as conduction loss.
Furthermore, they are largely incompatible with high-current inter-module bus-bar connections, e.g., in reconfigurable batteries \cite{bayati2025highly}.

This paper presents a solution that adds DM inductance around existing, shortest-path conductors. 
The proposed design does not wind the conductors. 
Instead, it winds the flux.

\section{Proposal of Figure-of-Eight Core}
\subsection{Core Shape and Magnetic Flux Path}

\begin{figure}[htbp]
    \centering
    \includegraphics[width=\columnwidth]{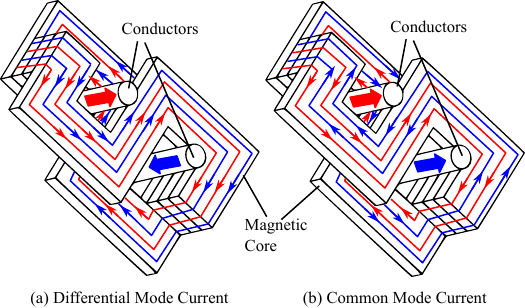}
    \caption{3D figure-of-eight-shaped magnetic core. Two parallel conductors pass through the two loops, one per loop. Under DM excitation, the flux reinforces throughout the entire magnetic path. Under CM excitation, it cancels.
    }
    \label{fig:core_illustration}
\end{figure}

The proposed design is a figure-of-eight-shaped magnetic core that routes the magnetic flux path around the existing conductors.
It consists of two loops joined at the two far ends but separated in the middle (Fig.~\ref{fig:core_illustration}).
Two shortest-path parallel conductors pass through the structure, one per loop.

Under CM excitation, which forms the modular multilevel or cascaded-bridge output, both conductors carry current in the same direction. 
Their magnetic fields cancel in the center section and thus the entire path, so the net inductance is negligible. 
Under DM excitation, which carries the local inter-module energy exchange, the currents flow in opposite directions. 
Their fields reinforce throughout the entire magnetic path and produce a significant inductance that opposes the DM current. 

\subsection{Scalable Implementation}
For large-quantity, series use, the core should obviously be sintered in one or two pieces that can be clipped on, similar to existing CM cores. For our prototype and small quantities, we decomposed the 3D structure into planar pieces that can be cut from flat ferrite sheets. 

\begin{figure*}[htbp]
    \centering
    \includegraphics{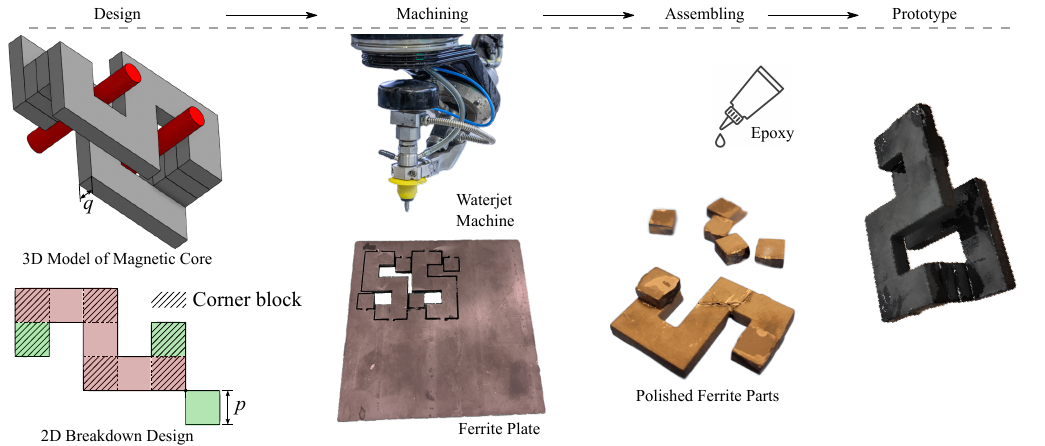}
    \caption{Prototype manufacturing. The 3D model is broken down to a 2D design for water-jet cutting of magnetic plates. The components are then bonded together with epoxy to obtain the figure-of-eight core. Red colors in the 2D breakdown indicate blocks in which the field direction is mostly parallel to the block, green colors indicate those perpendicular.
    }
    \label{fig:production_demonstration}
\end{figure*}

We designed two planar pieces: an S-shaped piece (\qty{42.3}{\milli\meter} $\times$ \qty{25.4}{\milli\meter} outer dimensions) and square end pieces (\qty{8.5}{\milli\meter} $\times$ \qty{8.5}{\milli\meter}). 
All pieces were cut from \qty{4}{\milli\meter} thick ferrite plates by water jet cutting (ProtoMAX Abrasive Waterjet, Omax Corporation, USA). 
Cutting multiple units simultaneously from a single plate keeps the per-unit cost low. 
Small tabs were added to the outline of each piece to keep it connected to the plate during cutting. 
After cutting, the pieces were deburred, polished with sandpaper, and bonded with epoxy. 
After a 24-hour cure, the assembly was complete. 
Fig.~\ref{fig:production_demonstration} shows the production sequence.

\subsection{Inductance and Loss Resistance Estimation}
The figure-of-eight core can be treated as a DM variant of typical clamp-on CM chokes, and its inductance follows:
\begin{equation}
    L_{\mathrm{DM}} =  L_{\mathrm{core}} + L_{\mathrm{leak}} =   \frac{N^2}{\mathcal{R}_{\mathrm{core}}} + L_{\mathrm{leak}},
\end{equation}
where $\mathcal{R}_{\mathrm{core}}$ is the core reluctance, $N$ the effective turn count, and $L_{\mathrm{leak}}$ the leakage inductance. 
Under DM excitation, the magneto-motive forces of the two conductors drive flux in the same direction and reinforce each other throughout the shared magnetic path, so their contributions add and $N=2$.

\subsubsection{Core geometry}
The ferrite core consists of square units of side length $p$ and thickness $q$, with relative permeability $\mu_{\mathrm{fe}}\gg 1$.
For the flux mostly traveling in the plane of each block, the path length is $p$ and the cross-sectional area is $p q$ (red blocks in Fig.~\ref{fig:production_demonstration}). Therefore, the reluctance is 
\begin{equation}
    \mathcal{R}_{\parallel} \approx \frac{p}{\mu_{\mathrm{fe}}\mu_0 p  q} = \frac{1}{\mu_{\mathrm{fe}}\mu_0 q}.
    \label{eq:R_parallel}
\end{equation}
For magnetic flux that travels through the thickness of each block, the path length is $q$ and the cross-sectional area is $p^2$ (green blocks in Fig.~\ref{fig:production_demonstration}). Therefore, the reluctance is 
\begin{equation}
    \mathcal{R}_{\perp} \approx \frac{q}{\mu_{\mathrm{fe}}\mu_0 p^2}.
    \label{eq:R_perp}
\end{equation}

For simplicity, we applied a correction factor of 0.5 to the reluctance of each corner block, simplified from the 0.56 corner correction in sheet resistance calculations for integrated circuits. The front and back faces of the figure-of-eight each contribute four corner blocks to $\mathcal{R}_{\parallel}$, and the top and bottom paths each contribute two corner blocks to $\mathcal{R}_{\perp}$ (hashed blocks in Fig.~\ref{fig:production_demonstration}).

\subsubsection{Gap Correction}
The epoxy joints between the ferrite pieces introduce gaps of sub-millimeter width $g$. Each gap has cross-sectional area $p^2$.
Despite their small size, these gaps dominate the reluctance due to their small relative permeability:
\begin{equation}
     \mathcal{R}_{\mathrm{gap}} \approx  \frac{g}{\mu_{\mathrm{gap}}\mu_{0} p^2} \approx \frac{g}{\mu_{0} p^2}.
    \label{eq:R_gap}
\end{equation}

\subsubsection{Combined Estimation}
With block numbers (see illustration in Fig.\ref{fig:production_demonstration}) of $n_{\parallel} = 2\cdot (n_{\parallel\mathrm{straight}}+n_{\parallel\mathrm{corner}}/2) = 10$, $n_{\perp}=2\cdot (n_{\perp\mathrm{straight}}+n_{\perp\mathrm{corner}}/2)=6$, and $n_{\mathrm{gap}}=6$, the total core reluctance is therefore
\begin{equation}
     \mathcal{R}_{\mathrm{core}} \approx n_{\parallel}\mathcal{R}_{\parallel} +  n_{\perp}\mathcal{R}_{\perp} + n_{\mathrm{gap}}\mathcal{R}_{\mathrm{gap}} 
    \label{eq:R_total}
\end{equation}
and the total inductance is 
\begin{equation}
    L_{\mathrm{core}} = \frac{2\mu_{\mathrm{fe}}\mu_0 q}{5+3(\frac{q}{p})^2+3\mu_{\mathrm{fe}}g\frac{q}{p^2}}.
    \label{eq:L_final}
\end{equation}

\subsubsection{Loss Resistance}
The ferrite material has a loss tangent $\tan \delta_{\mu}$ between \qty{0.02}{} and \qty{0.03}{} for frequencies below \qty{300}{kHz}. The loss resistance of the core in the linear operation range can therefore be given as $R_{\mathrm{DM}} = 2\pi f L_{\mathrm{DM}} \tan \delta_{\mu}$.

\section{Results}
\subsection{Simulation}
We simulated the magnetic field distribution in COMSOL Multiphysics v6.3.
The simulation performed \textit{Stationary} studies using the \textit{Magnetic Fields (mf)} interface in the AC/DC module.

\begin{figure*}[htbp]
    \centering
    \includegraphics[width=0.90\textwidth]{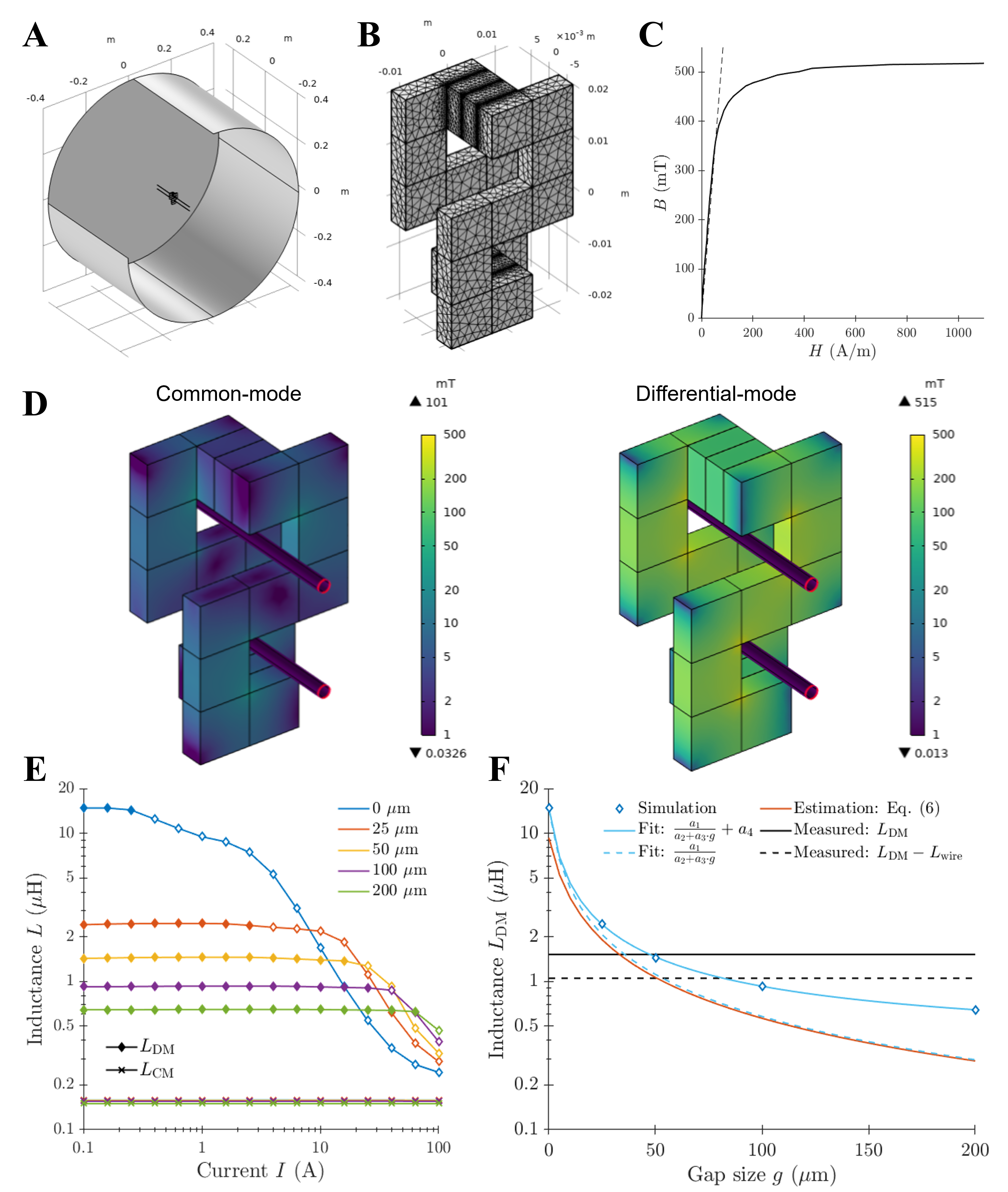}
    \caption{The modeling setup in COMSOL, showing the cylindrical simulation space (A), details of meshing (physics-controlled \textit{Fine} setting) revealing narrow gaps of epoxy/air (B), and the magnetization curve of the ferrite of the magnetic core (C). Simulated magnetic field distribution (D) for CM (left) and DM (right), shown for \qty{10}{A} current and epoxy gaps of \qty{50}{\micro\meter} width each and with the same logarithmic color scale. Inductance (E) of CM (crosses) and DM (diamonds) simulation as a function of current, with DM inductance that has saturation, i.e., inductance reduced to below 95\% of low-current values, shown with open markers. DM inductance versus gap size (F), showing simulation (diamonds), least-square fit including and excluding a term for wire inductance (blue solid and dashed lines), and simplified estimation Eq.\ref{eq:L_final} of core inductance only (orange line), with measurements of total DM and DM core inductance without wires (black solid and dashed lines).}
    \label{fig:simulation_all}
\end{figure*}

The geometry reproduced the fabricated core at full scale.
The simulation domain was a cylinder of \qty{0.4}{m} radius and \qty{0.5}{m} length, centered on the ferrite core, to capture the fringing flux (Fig.~\ref{fig:simulation_all}A).
The epoxy joints were modeled as air gaps of equal thickness with five values tested: \qty{0}{}, \qty{25}{}, \qty{50}{}, \qty{100}{}, and \qty{200}{\micro\meter} (Fig.~\ref{fig:simulation_all}B).

The ferrite regions were assigned the $B$--$H$ curve of the BH1T ferrite (FPL Series, KEMET Electronics). 
This material has a relative permeability of $\mu_r \approx 5300$ (estimated from the $B$--$H$ curve; nominal value $3000\pm 25\%$) for fields below approximately \qty{50}{A/m} or \qty{330}{mT}. It saturates at approximately \qty{520}{mT} (Fig.~\ref{fig:simulation_all}C).
The two conductors were modeled as solid copper cylinders of \qty{20}{cm} length and \qty{2.0}{\milli\meter} diameter, equivalent to American Wire Gauge (AWG) 12 wires. The current in each wire was specified as a surface current density, corresponding to a total current $I$ that was swept between \qty{0.1}{A} and \qty{100}{A}, with five values per decade and a total of 16 current levels. 
The currents were in the same direction for CM excitation and opposite for DM excitation. 
The outer boundary of the air domain was set as magnetic insulation ($\mathbf{n} \times \mathbf{A} = 0$).

We extracted the secant inductance for each excitation mode from the magnetic energy in the entire space $U$ (COMSOL global evaluation \textit{mf.intWm}) as $L=2U/I^2$. The simulation verified magnetic flux cancellation under CM excitation and flux reinforcement under DM excitation (Fig.~\ref{fig:simulation_all}D). The CM inductance was low (all between \qty{0.15}{\micro\henry} and \qty{0.16}{\micro\henry}), with minimal differences for different thicknesses of epoxy gaps and current levels (Fig.~\ref{fig:simulation_all}E). In contrast, the DM inductance was much higher and varied with the thickness of the epoxy gap. With no epoxy gap, the DM inductance $L_{\mathrm{DM}}$ was highest at low current levels (\qty{14.8}{\micro\henry} for $I\leq 0.25 \ \mathrm{A}$), but dropped as the core saturated with larger currents. 
With flux-limiting epoxy gaps, the DM inductance present was lower (between \qty{0.59}{\micro\henry} and \qty{2.45}{\micro\henry} for different gap sizes) but stable across a wide range of current amplitudes. Saturation currents increased with the gap size, and mostly did not occur until $I\geq 25 \ \mathrm{A}$, except for \qty{25}{\micro\meter} gaps with saturation at \qty{4}{\ampere}.

The calculated inductance values $L_{\mathrm{core}}$ were within a factor of two of the simulation results for the same gap size (Fig.~\ref{fig:simulation_all}F). The primary sources of discrepancy are the simplified corner reluctance correction, as well as fringing flux at the gap faces and the leakage inductance not captured by the analytical model. The theoretical analysis thus provided reasonably good estimates of the ferrite-associated incremental inductance despite these simplified assumptions. The simulated DM inductance matched Eq.~\ref{eq:L_final} with a constant term \qty{0.35}{\micro\henry} representing the leakage inductance; the calibrated curve without this term matched the theoretical estimation very well for gap widths larger than \qty{50}{\micro\meter}.
The simulated inductance for epoxy gaps of \qty{50}{\micro\meter} width each (\qty{300}{\micro\meter} total), \qty{1.45}{\micro\henry} (or \qty{1.10}{\micro\henry} after subtracting the fitted constant term), best matched the measurements (see next section). 

\subsection{Experimental Validation}

\begin{figure}[htbp]
    \centering
    \includegraphics[width=0.9\columnwidth]{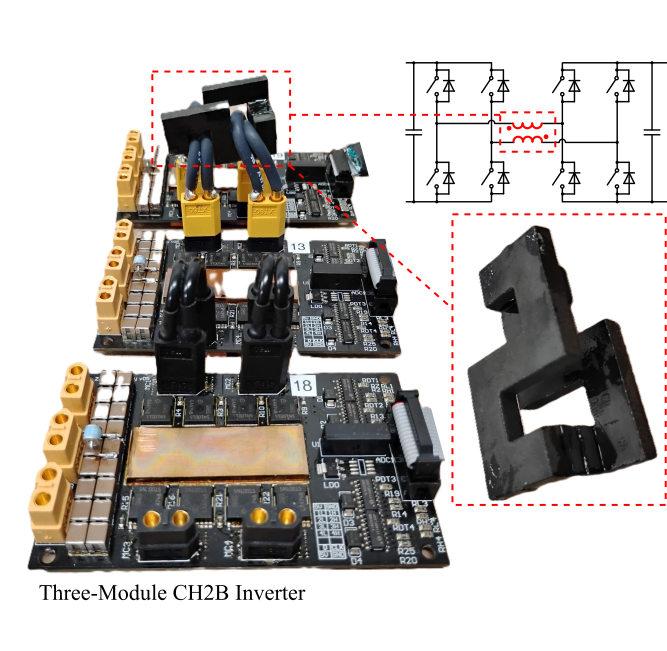}
    \caption{Experimental platform boards used for testing.}
    \label{fig:platform_boards}
\end{figure}

\begin{table}[htbp]
    \centering
    \caption{Core Parameters and Measured Results}
    \label{tab:combined_params}
    \renewcommand{\arraystretch}{1.25}
    \begin{tabular}{m{0.25\columnwidth}m{0.30\columnwidth}>{\centering\arraybackslash}m{0.30\columnwidth}}
        \toprule
        \textbf{Category} & \textbf{Parameter} & \textbf{Value} \\
        \midrule
        \multirow{2}{0.25\columnwidth}{\raggedright Estimation \newline (\qty{300}{\micro\meter} total gap)}
         & DM inductance & 1.06~\unit{\micro\henry} \\
         & DM loss resistance & 2.00~\unit{\milli\ohm} (\qty{10}{kHz})\\
        \midrule
        \multirow{3}{0.25\columnwidth}{\raggedright Simulation \newline (\qty{300}{\micro\meter} total gap)}
         & CM inductance & 0.16~\unit{\micro\henry} \\
         & DM inductance & 1.45~\unit{\micro\henry} \\
         & DM saturation current & 25~\unit{\ampere} \\
        \midrule
        \multirow{3}{0.25\columnwidth}{\raggedright Measurement \newline (\qty{10}{kHz})}
         & Single conductor & \qty{1.2}{\milli\ohm},\ \qty{0.47}{\micro\henry} \\
         & CM impedance & \qty{0.6}{\milli\ohm},\ \qty{0.11}{\micro\henry} \\
         & DM impedance & \qty{2.4}{\milli\ohm},\ \qty{1.52}{\micro\henry} \\
        \bottomrule
    \end{tabular}
\end{table}
We built a three-module series\,/\,parallel modular multilevel inverter based on the cascaded double H-bridge (CH2B) topology.
Fig.~\ref{fig:platform_boards} shows the experimental platform.
For maximum DM load, a 5~V DC source supplies the middle module so that all modules receive their voltage at 5~V through switched-capacitor power transfer with parallelization.
The load is resistive. The inverter produces a \qty{50}{\hertz} sinusoidal output.
Nearest-level modulation serves to simplify control.
Even with a low CM load ($\sim$10~mA), a module voltage fluctuation below \qty{0.5}{\volt} drives the inter-module current surge above 20~\unit{\ampere}.

We installed the fabricated core on the same three-module inverter platform to increase the inter-module DM impedance.
It added no extra conductors or resistance.
We measured impedance using a bench LCR\,/\,ESR meter at \qty{10}{kHz} (Model 889A, B\&K Precision Corporation): the bare interconnection wire alone, and the interconnection under CM and DM excitation with the figure-of-eight core installed (Table~\ref{tab:combined_params}). 
The measured DM inductance of \qty{1.52}{\micro\henry} agreed well with the simulation result of \qty{1.45}{\micro\henry} and, after subtracting the \qty{0.47}{\micro\henry} inductance of the wires, also matched the analytical estimate $L_{\mathrm{core}}$ of \qty{1.06}{\micro\henry} and the simulated \qty{1.10}{\micro\henry} without the fitted constant term (Fig.~\ref{fig:simulation_all}F). The DM resistance also matched the estimated loss resistance (Table~\ref{tab:combined_params}).

\begin{figure}[htbp]
    \centering
    \includegraphics[width=0.9\columnwidth]{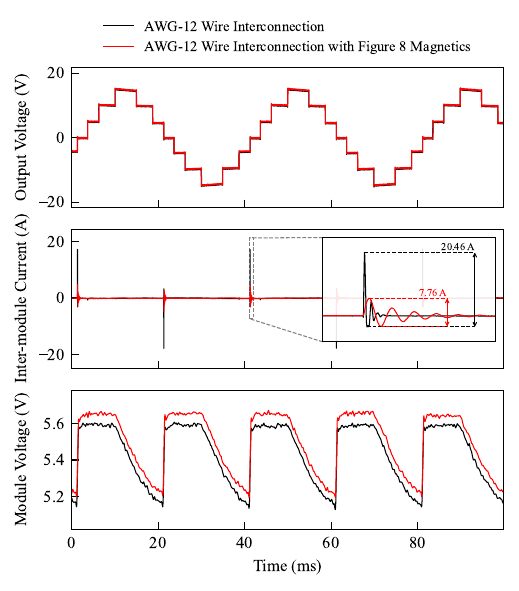}
    \caption{Measured waveforms of current surge with and without the clip-on figure-of-eight core.}
    \label{fig:current_surge_validation_figure8core}
\end{figure}

With the system operated under the same control scheme and input\,/\,output configurations, the output voltage with and without the figure-of-eight core agreed well and showed similar module voltage fluctuations (Fig.~\ref{fig:current_surge_validation_figure8core}). The figure-of-eight core, however, reduced and spread out the current surge during switching transients from 20.46~\unit{\ampere} to 7.76~\unit{\ampere}.

\section{Conclusion}
This paper presented a highly compact resistance-free solution for DM inductance. 
A figure-of-eight-shaped magnetic core routes the flux path around existing conductors.
This design requires no conductor re-routing and has low CM inductance due to magnetic flux cancellation in the core.
The compact, retrofit-friendly design can be fabricated from flat ferrite sheets by water jet cutting. 
The measured DM inductance is \qty{1.52}{\micro\henry}, and the added resistance is negligible.
By contrast, the CM inductance is suppressed to only 7\% of the DM inductance.
The experimental results demonstrate a 62.1\% reduction in the inter-module current surge. 

\bibliographystyle{Bibliography/IEEEtranTIE}
\bibliography{Bibliography/IEEEabrv,Bibliography/main}\ 

\end{document}